\documentclass[aps,prl,twocolumn,floatfix,superscriptaddress,showpacs]{revtex4-1}
\usepackage{graphicx}
\usepackage[pdftex,colorlinks,citecolor=blue,bookmarks]{hyperref}
\begin{document}
\title{$Q$ value and half-life of double-electron capture in $^{184}$Os}%

\author{C.~Smorra}
\affiliation{Max-Planck-Institut f\"ur Kernphysik, Saupfercheckweg 1, D-69117 Heidelberg} 
\affiliation{Institut f\"ur Kernchemie, Johannes Gutenberg-Universit\"at, Fritz-Strassmann-Weg 2, D-55128 Mainz} 
\affiliation{Fakult\"at f\"ur Physik und Astronomie, Ruprecht-Karls-Universit\"at, Philosophenweg 12, D-69120 Heidelberg} 
\author{T.~R.~Rodr\'iguez}
\affiliation{Technische Universit\"at Darmstadt, Institut f\"ur Kernphysik, Schlossgartenstr. 2, D-64289 Darmstadt, Germany} 
\affiliation{GSI Helmholtzzentrum f\"ur Schwerionenforschung, Planckstra\ss e 1, D-64291 Darmstadt} 
\author{T.~Beyer}
\affiliation{Max-Planck-Institut f\"ur Kernphysik, Saupfercheckweg 1, D-69117 Heidelberg} 
\affiliation{Fakult\"at f\"ur Physik und Astronomie, Ruprecht-Karls-Universit\"at, Philosophenweg 12, D-69120 Heidelberg} 
\author{K.~Blaum}
\affiliation{Max-Planck-Institut f\"ur Kernphysik, Saupfercheckweg 1, D-69117 Heidelberg} 
\affiliation{Fakult\"at f\"ur Physik und Astronomie, Ruprecht-Karls-Universit\"at, Philosophenweg 12, D-69120 Heidelberg} 
\author{M.~Block}
\affiliation{GSI Helmholtzzentrum f\"ur Schwerionenforschung, Planckstra\ss e 1, D-64291 Darmstadt} 
\author{Ch.E.~D\"ullmann}
\affiliation{Institut f\"ur Kernchemie, Johannes Gutenberg-Universit\"at, Fritz-Strassmann-Weg 2, D-55128 Mainz} 
\affiliation{GSI Helmholtzzentrum f\"ur Schwerionenforschung, Planckstra\ss e 1, D-64291 Darmstadt} 
\affiliation{Helmholtz-Institut Mainz, Johann-Joachim-Becher-Weg 36, D-55128 Mainz} 
\author{K.~Eberhardt}
\affiliation{Institut f\"ur Kernchemie, Johannes Gutenberg-Universit\"at, Fritz-Strassmann-Weg 2, D-55128 Mainz} 
\affiliation{Helmholtz-Institut Mainz, Johann-Joachim-Becher-Weg 36, D-55128 Mainz}  
\author{M.~Eibach}
\affiliation{Institut f\"ur Kernchemie, Johannes Gutenberg-Universit\"at, Fritz-Strassmann-Weg 2, D-55128 Mainz} 
\affiliation{Fakult\"at f\"ur Physik und Astronomie, Ruprecht-Karls-Universit\"at, Philosophenweg 12, D-69120 Heidelberg} 
\author{S.~Eliseev}
\affiliation{Max-Planck-Institut f\"ur Kernphysik, Saupfercheckweg 1, D-69117 Heidelberg} 
\author{K.~Langanke}
\affiliation{Technische Universit\"at Darmstadt, Institut f\"ur Kernphysik, Schlossgartenstr. 2, D-64289 Darmstadt, Germany} 
\affiliation{GSI Helmholtzzentrum f\"ur Schwerionenforschung, Planckstra\ss e 1, D-64291 Darmstadt} 
\affiliation{Frankfurt Institute for Advanced Studies, Frankfurt, Ruth-Moufang Str. 1, D-60438 Frankfurt, Germany} 
\author{G.~Mart\'inez-Pinedo}
\affiliation{Technische Universit\"at Darmstadt, Institut f\"ur Kernphysik, Schlossgartenstr. 2, D-64289 Darmstadt, Germany} 
\affiliation{GSI Helmholtzzentrum f\"ur Schwerionenforschung, Planckstra\ss e 1, D-64291 Darmstadt} 
\author{Sz.~Nagy}
\affiliation{Max-Planck-Institut f\"ur Kernphysik, Saupfercheckweg 1, D-69117 Heidelberg} 
\affiliation{GSI Helmholtzzentrum f\"ur Schwerionenforschung, Planckstra\ss e 1, D-64291 Darmstadt} 
\author{W.~N\"ortersh\"auser}
\affiliation{Institut f\"ur Kernchemie, Johannes Gutenberg-Universit\"at, Fritz-Strassmann-Weg 2, D-55128 Mainz} 
\affiliation{GSI Helmholtzzentrum f\"ur Schwerionenforschung, Planckstra\ss e 1, D-64291 Darmstadt} 
\author{D.~Renisch}
\affiliation{Max-Planck-Institut f\"ur Kernphysik, Saupfercheckweg 1, D-69117 Heidelberg}
\affiliation{Institut f\"ur Kernchemie, Johannes Gutenberg-Universit\"at, Fritz-Strassmann-Weg 2, D-55128 Mainz}
\author{V.M.~Shabaev}
\affiliation{Department of Physics, St.~Petersburg State University, 198504 St.~Petersburg, Russia}
\author{I.I.~Tupitsyn}
\affiliation{Department of Physics, St.~Petersburg State University, 198504 St.~Petersburg, Russia}
\author{N.A.~Zubova}
\affiliation{Department of Physics, St.~Petersburg State University, 198504 St.~Petersburg, Russia}

\date{\today} 

\begin{abstract}
$^{184}$Os has been excluded as a promising candidate for the search of neutrinoless double-electron capture. High-precision mass measurements with the Penning-trap mass spectrometer TRIGA-TRAP result in a marginal resonant enhancement with $\Delta$ = -8.89(58)~keV excess energy to the 1322.152(22)~keV  $0^+$ excited state in $^{184}$W. State-of-the-art energy density functional calculations are applied for the evaluation of the nuclear matrix elements to the excited states predicting a strong suppression due to the large deformation of mother and daughter states. The half-life of the transition exceeds $T_{1/2}  (^{184}\mathrm{Os}) \geq 1.3 \times 10^{29}$~y for an effective  neutrino mass of 1~eV.
\end{abstract}

\maketitle

The observation of neutrinoless double-beta transitions would reveal
physics beyond the Standard Model, as it would establish neutrinos to
be Majorana particles, which implies a violation of the lepton number
conservation. Experiments searching for these transitions have focused
on the detection of neutrinoless double-beta decay (0$\nu\beta\beta$)
rather than neutrinoless double-electron capture
(0$\nu\epsilon\epsilon$). One reason among others is in general the significantly shorter
half-life of the 0$\nu\beta\beta$ process. However, in the
case of neutrinoless double-electron capture, the transition is
expected to be resonantly enhanced if the initial and the final state
of the transition are degenerate in energy \cite{Voloshin1982,
  Bernabeu198315,Krivoruchenko2011140}.

In this work, we investigate neutrinoless double-electron capture in $^{184}$Os
to various nuclear states. In particular, the transition to the
nuclear excited 0$^+$ state with an energy of 1322.152(22) keV by a capture
of two K-electrons was tagged as partially resonantly enhanced~\cite{Krivoruchenko2011140}. The
capture of two K-electrons, the low spin of the nuclear excited state, and
the large nuclear radius could result in a reasonably short half-life
compared to other transitions even without being fully resonantly
enhanced~\cite{Krivoruchenko2011140}.

The capture rate of a 0$\nu\epsilon\epsilon$ transition is given
by~\cite{Bernabeu198315, Krivoruchenko2011140} 
\begin{equation}
\lambda_{\epsilon\epsilon} =
\frac{(\lambda G_F \left|V_{ud}\right|)^4}{(4\pi R)^2} |m_{\epsilon\epsilon}|^2
\left|M^{\epsilon\epsilon}\right|^2
P_{\epsilon\epsilon} \left(\frac{\Gamma_{2h}}{\Delta^2
    + \Gamma_{2h}^2 / 4}\right),
\label{EQ_1}
\end{equation}
where $\lambda=g_A/g_V=1.2701(25)$ \cite{PDG} is the ratio of the axial vector and the 
vector coupling constant, $G_F$ the Fermi coupling constant, $\left|V_{ud}\right|$ the $ud$ element of the Cabibbo-Kobayashi-Maskawa (CKM) matrix, $R$ the
nuclear radius, $|m_{\epsilon\epsilon}|$ the effective electron
neutrino mass, $|M^{\epsilon\epsilon}|$ the nuclear matrix element
(NME) of the transition that consists dominantly of Fermi and
Gamow-Teller components~\cite{TRR_GMP_ECEC}, $P_{\epsilon\epsilon} =
\left|\Psi_{h1}\right|^2 \left|\Psi_{h2}\right|^2$ the absolute square
of the two electron wave functions at the nucleus, $\Gamma_{2h}$ the width of the two-electron hole state in the daughter nucleus, and
$\Delta$ the excess energy.  In order to estimate the capture rate,
we have performed calculations of the NME by state-of-the-art energy
density functional (EDF) calculations. These methods have been
successfully applied to nuclear structure calculations (see
ref.~\cite{Bender.Heenen.Reinhard:2003} for a review) and they have
been recently implemented in calculations of neutrinoless double-beta
decay and double-electron capture
NMEs~\cite{PRL_105_252503_2010,TRR_GMP_ECEC}. In order to derive
$\Delta=Q_{\epsilon\epsilon} - E_\gamma - B_{2h}$, we have
experimentally determined the double-electron capture $Q$ value
$Q_{\epsilon\epsilon}$, which is the difference of the atomic masses
of the mother and daughter nuclides. $E_\gamma$ denotes the
nuclear excitation energy of the final state, and $B_{2h}$ is the
excitation energy of the two-electron hole state in the daughter
nucleus. Typically, the uncertainty of $\Delta$ is limited by the uncertainty of $Q_{\epsilon\epsilon}$. In the case of double-electron
capture in $^{184}$Os, the $Q$ value of $Q_{\epsilon\epsilon} =$ 1451.2(1.6) keV \cite{Audi2003337} was only obtained indirectly from the masses of the
other Os and W isotopes via (n,$\gamma$)-reactions. $Q$ values taken from mass values in Ref.~\cite{Audi2003337} can be rather inaccurate as the measurements in the references \cite{Kolhinen2011116, PhysRevC.84.028501} have revealed. Thus, for all potentially resonantly
enhanced transitions, their $Q$ values must be measured directly with Penning traps, which are presently the tool of choice \cite{Klaus20061}.

The resonance condition of several promising transitions have recently been investigated with Penning traps \cite{PhysRevC.84.012501, PhysRevLett.106.052504, PhysRevC.81.032501, Kolhinen2011116, PhysRevC.84.028501, PhysRevLett.107.152501, Smorra:12:1}, and a strong resonant enhancement was found for two nuclides, $^{152}$Gd \cite{PhysRevC.84.012501} and $^{156}$Dy \cite{PhysRevLett.106.052504}.

The half-life for double-electron capture in $^{184}$Os has recently
been estimated in ref.~\cite{Krivoruchenko2011140} using an empirical
value for the NME, which corresponds to the maximum value obtained for
medium-mass nuclei assuming spherical shapes for initial and final
states~\cite{Simkovic.Faessler.ea:2009}. However, mid-shell nuclei like
$^{184}$Os and $^{184}$W are known to be significantly
deformed~\cite{NuDat2}. As NMEs have been shown to be very sensitive
to deformation~\cite{PRL_105_252503_2010,TRR_GMP_ECEC}, we have
calculated the NME using a method that consistently describes the
deformation of the nuclear states involved. We have further
recalculated the electron wave functions and atomic excitation
energies in the daughter nuclide. Hence, we report about improvements
of three crucial input parameters to the resonance enhancement factor, which
allows us to check if the $^{184}$Os and $^{184}$W pair is a viable
candidate to observe neutrinoless double-electron capture.

The $Q$-value measurement reported here was performed with the double-Penning-trap mass spectrometer TRIGA-TRAP \cite{Ketelaer2008162}. The
$Q$ value is determined by measuring the ratio $r$ of the cyclotron
frequencies $\nu_c = q B /( 2 \pi M)$ of $^{184}$Os$^+$ to $^{184}$W$^+$
ions stored in a Penning trap with a magnetic field $B$:
\begin{equation}
Q_{\epsilon\epsilon} / c^2 = \left(M(^{184}\mathrm{Os}) - m_e\right) \left( 1 - \frac{\nu_c(^{184}\mathrm{Os}^+)}{\nu_c(^{184}\mathrm{W}^+)} \right),
\label{EQ_3}
\end{equation}
where $m_e$ denotes the electron mass. This method provides a direct
measurement of the $Q$ value. Note that the absolute atomic masses
$M$ of these two nuclides cannot be derived from such a
measurement. However, the absolute masses of the two nuclides were
determined within this work by measuring the ratio $r' =
\nu_c(\mathrm{C}_{15}^+) / \nu_c$ of the cyclotron frequencies of the
carbon cluster ion $^{12}$C$_{15}^+$ and the investigated nuclide, in
order to obtain the mass values via $M = (M_{\mathrm{C}_{15}} - m_e) r' + m_e$.

\begin{figure}[htb]
\centering
\includegraphics[width=\linewidth]{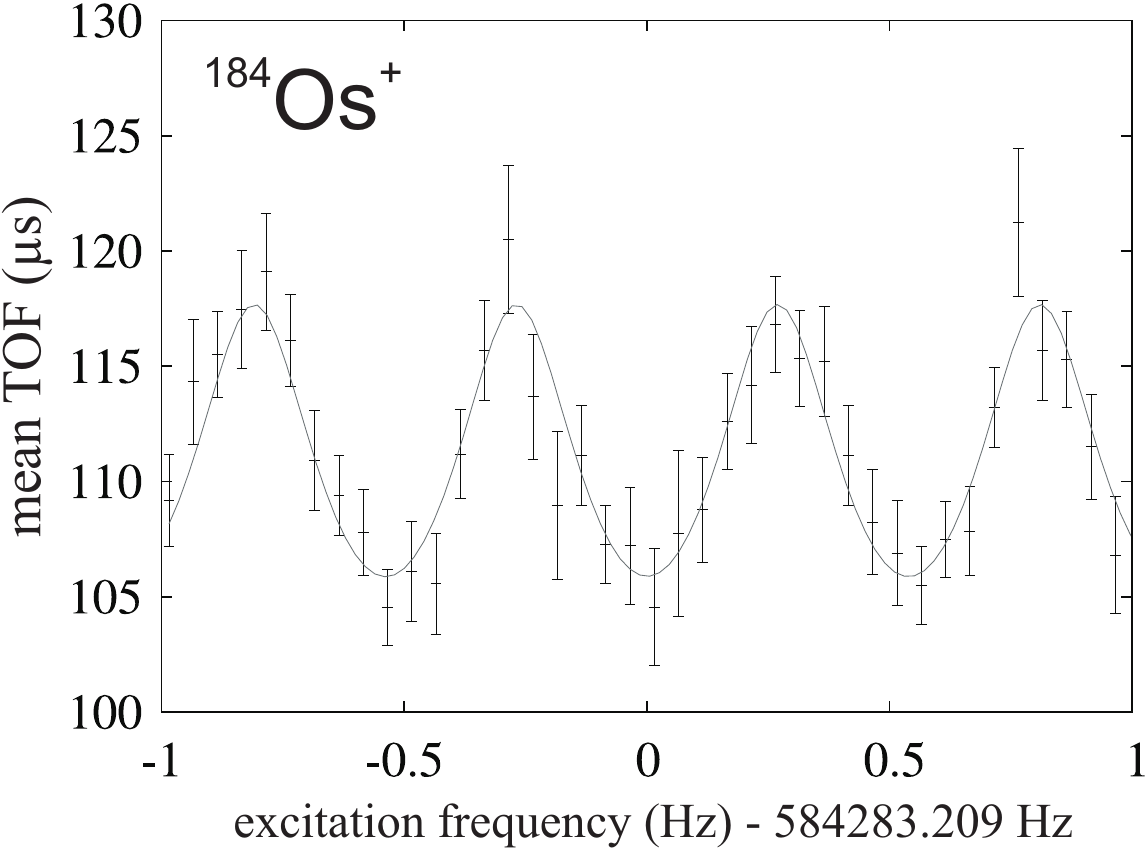}
\caption[Osmium-184 resonance]{Time-of-flight ion cyclotron resonance of $^{184}$Os$^+$ with about 450 ions. The data points show the mean time-of-flight values as function of the excitation frequency. The solid line is a fit of the theoretical line shape \cite{Martin2007122} to the data points. }
\label{FIG_2}
\end{figure}

A laser ablation ion source as described in
ref.~\cite{0953-4075-42-15-154028} was used to produce $^{184}$Os and
$^{184}$W ions from solid samples of osmium and tungsten. Due to the
low natural abundance of $^{184}$Os (0.02$\%$), an enriched sample with
1.5$\%$ abundance of $^{184}$Os was used for the ion production.
The ion pulses from the ion source were captured in the cylindrical Penning trap, called purification trap. There, the
captured ion pulse was cleaned from unwanted ion species using a
mass-selective buffer-gas cooling technique \cite{Savard1991247} with
 mass resolving power of about 3$\times 10^4$. By using two different
targets for $^{184}$W and $^{184}$Os, the presence of isobaric
contaminants in the purification trap was avoided. A cooled monoisotopic ion bunch was transferred into the hyperbolical Penning
trap, called precision trap, where the actual mass measurement took
place. The cyclotron frequency of the stored ions was measured via the
time-of-flight ion-cyclotron-resonance (TOF-ICR) method with a Ramsey
excitation scheme \cite{George2007110} using two excitation pulses of
200 ms and a waiting time of 1600 ms in between. A typical cyclotron
resonance for $^{184}$Os$^+$ is shown in Fig.~\ref{FIG_2}. The
cyclotron frequency has been obtained from a fit of the
theoretical line shape \cite{Martin2007122} to the experimental
data.

\begin{table}[b!]
\centering
\caption{Results of the mass measurements. For each ion of interest the reference ion species is $^{12}$C$^{+}_{15}$. The cyclotron-frequency ratio $r'$, the mass excess ME = $M$ - $A$ u, where u is the unified atomic mass unit, and the mass excess ME$_{\mathrm{lit}}$ listed in Ref.~\cite{Audi2003337} are given.}
\begin{tabular}{cccccc} \hline \hline
Ion & $r'$ & ME (keV) & ME$_{\mathrm{lit}}$ (keV) \\ \hline
$^{184}$Os$^+$ &1.0219583764(67) & -44251.47(1.13) & -44256.60 (1.34) \\ 

$^{184}$W$^+$  & 1.0219496960(55) & -45705.40(0.94) & -45707.54 (0.87) \\ 

$^{12}$C$_{16}^+$ & 1.0666668659(70) & -0.62(1.25)  & 0 (0) \\ \hline \hline
\end{tabular}
\label{TABLE_2}
\end{table}

\begin{figure}[htb]
\centering
\includegraphics[width=\linewidth]{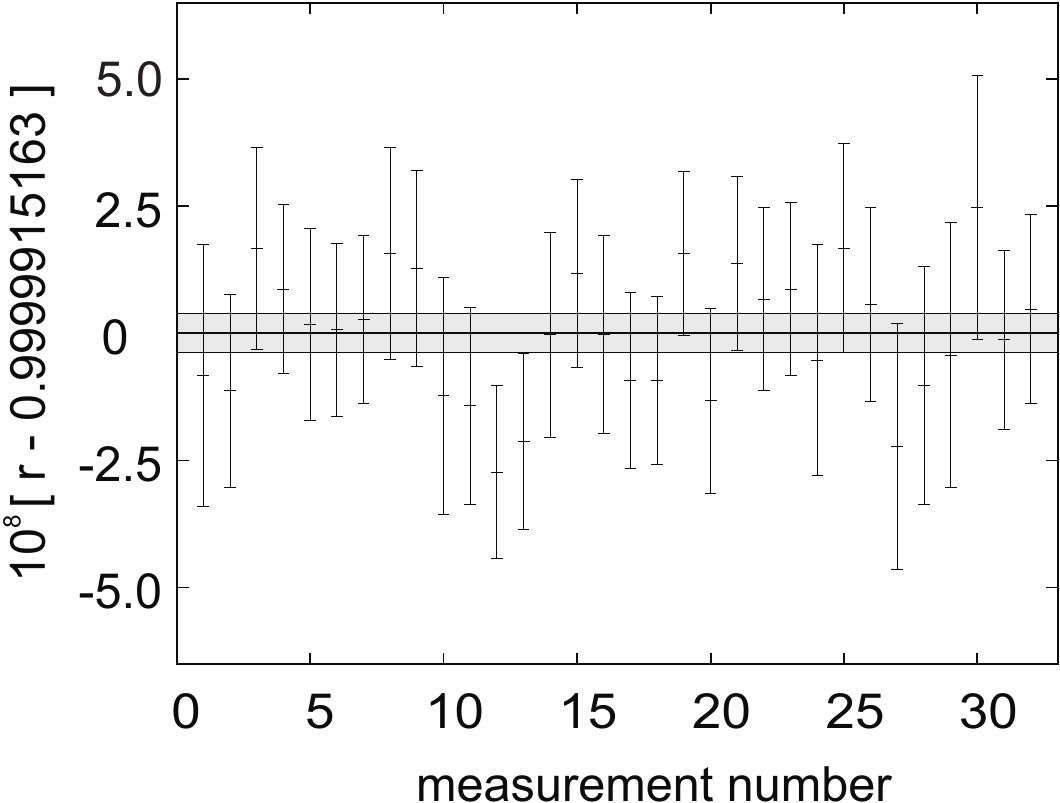}
\caption[Osmium-184 frequency ratios]{The data points show the 32 cyclotron-frequency ratio  measurements recorded to determine the double-electron capture $Q$ value of $^{184}$Os. The thick line represents the weighted-mean value and the gray area its 1 $\sigma$-uncertainty.}
\label{FIG_3}
\end{figure}

In order to determine the $Q$ value according to Eq.~(\ref{EQ_3}), the
cyclotron-frequency ratio $r$ of $^{184}$Os$^+$ to $^{184}$W$^+$ was
determined in alternating measurements of the individual cyclotron
frequencies (see Fig.~\ref{FIG_3}). In the evaluation
procedure for cyclotron frequency ratios, systematic uncertainties as
described in detail in
ref.~\cite{springerlink:10.1140/epjd/e2010-00092-9} were
considered. The systematic shift of the frequency ratio depending on
the mass difference $\Delta m$ between the two ion species used in the
measurement is negligible ($\varepsilon_m/r < 10^{-11}$) in case of
nuclides with the same mass number $A$, as in a $Q$-value measurement
of a 0$\nu\epsilon\epsilon$ process. For the measurement of the
individual masses, this mass-dependent effect with a magnitude of
$\varepsilon_m = 2.2(0.2) \times 10^{-9} \Delta m$/u was considered. It
arises from a possible slight misalignment between the electric and
magnetic fields of the Penning trap
\cite{springerlink:10.1140/epjd/e2010-00092-9}.

The result of the cyclotron-frequency ratio measurement of
$^{184}$Os$^+$ to $^{184}$W$^+$ and the $Q$ value derived with
Eq.~(\ref{EQ_3}) are $0.9999915163(38)$ and $1453.68(0.58)$ keV,
respectively. The uncertainty of the literature $Q$ value was improved
by a factor of 3. A slight deviation below 1.5 $\sigma$ from our result to the value in Ref.~\cite{Audi2003337} ($Q_{\epsilon\epsilon}$ = 1451.2(1.6) keV) was found.

The atomic mass excess values obtained from the cyclotron-frequency ratio measurements in this work are listed in Tab.~\ref{TABLE_2}. Our measurement is a direct comparison to the atomic mass standard, unlike the values from Ref.~\cite{Audi2003337}, which show a deviation of 5.13(1.75) keV and 2.14(1.28) keV for $^{184}$Os and $^{184}$W, respectively. The $Q$ value extracted from our values $\mathrm{ME}(^{184}$Os$)$ and $\mathrm{ME}(^{184}$W) listed in Tab.~\ref{TABLE_2} is 1453.93(1.47)~keV and agrees well with our direct $Q$-value measurement. 


In order to determine the excess energy $\Delta$ to the excited 0$^+$
state with 1322.152(22) keV energy in $^{184}$W \cite{Baglin2010},
the excitation energy of the two-electron hole state $B_{\mathrm{2h}}$ and its width $\Gamma_{\mathrm{2h}}$
have been calculated. The value of $B_{\mathrm{2h}}$ was obtained by
applying the Dirac-Fock method \cite{Droese:2012:Ref28}, where
frequency-dependent Breit interaction, quantum electrodynamics (QED)
and correlation corrections were included. The calculations were
performed for the Fermi model of the nuclear charge distribution (with
$r_{\mathrm{RMS}}$=5.3670 fm for $^{184}$W \cite{ADNDT_84_185_2004}),
resulting in $B_{\mathrm{2h}}$= 140.418(12) keV. 
The value for the width of the autoionizing state of $^{184}$W with two
K-shell holes $\Gamma_{\mathrm{2h}}$ is approximated by using the double
value of the width of the single K-level hole \cite{Krivoruchenko2011140}.
The single K-level hole width is taken from Ref.~\cite{Campbell}. 
We estimate the uncertainty of $\Gamma_{\mathrm{2h}}$ by using the most conservative value
in Ref.~\cite{Campbell} and increase it by 2 eV to account for the uncertainty due to the hole-correlation effect suppressed approximately by a factor 1/Z. The contribution of the width of the excited nuclear state has been neglected. This approach resulted in $\Gamma_{\mathrm{2h}} = 80(10)$ eV. Prior to this work, $\Delta$ was $-11.3(1.6)$ keV using Ref.~\cite{Audi2003337} for the two mass values. The uncertainty given is obtained by adding the individual uncertainties of 1.3 keV and 0.9 keV of $^{184}$Os and $^{184}$W, respectively. Using our measured value, $\Delta$ becomes $-8.89(58)$ keV.

\begin{table*}[htb]
  \centering
  \begin{tabular}{lcccccccccc}
    \hline\hline
    & $E(0^{+}_{2})$ &  $E(0^{+}_{3})$ &  $E(2^+_1)[^{184}$Os]&  $E(2^+_1)[^{184}$W] & $B(E2)[^{184}$Os] & $B(E2)[^{184}$W] & $M^{\epsilon\epsilon}(0^{+}_{1})$ &
    $M^{\epsilon\epsilon}(0^{+}_{2})$ & 
    $M^{\epsilon\epsilon}(0^{+}_{3})$ & 
    $M^{\epsilon\epsilon}(\mathrm{Sph.})$ \\ \hline\hline
    Expt. ~\cite{ENSDF} & 1.002 & \textbf{1.322}  & 0.120 & 0.111 &
    99.6 & 119.8 & -- & -- & -- & -- \\
    EDF D1M & 1.135 & 3.169 & 0.172 & 0.155 & 135.0 & 146.6 & 0.631
    & 0.504 & 0.163 & 8.177 \\
    \hline\hline
  \end{tabular}
  \caption{Excitation energies (in MeV)  of the first two $0^{+}$
    states for $^{184}$W and   of the $2^+_1$ state and its quadrupole
    transition probability (in Weisskopf units) for $^{184}$Os and
    $^{184}$W are compared with EDF calculations based on the Gogny D1M functional.  The values of the NMEs,
    $M^{\epsilon\epsilon}$, to the three $0^+$ states in $^{184}$W and
    the NME obtained assuming spherical shapes ($\beta=0$) are also
    included. The experimental resonant state is written in
    boldface.\label{tableEDF2}} 
\end{table*}

\begin{figure}[htb]
\centering
\includegraphics[scale=0.7]{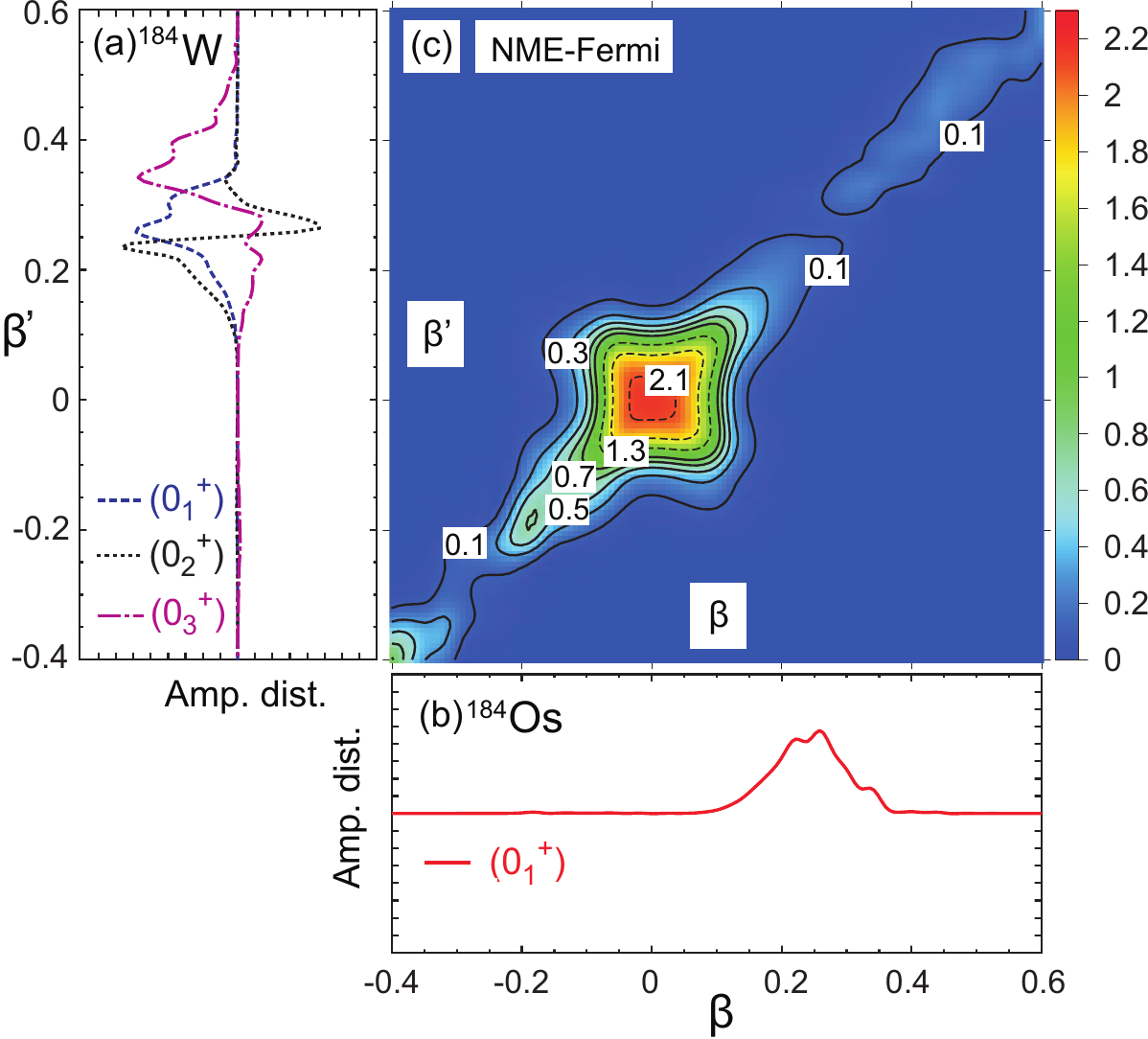}
\caption{(Color online) Deformation amplitudes for the three first $0^{+}$ excited states of $^{184}$W
  (a) and the ground state of $^{184}$Os (b) calculated with EDF Gogny D1M. (c) Intensity of the Fermi component of the $0\nu\epsilon\epsilon$ NME as a function of
  the quadrupole deformations $\beta$ and $\beta'$ of the $^{184}$Os and $^{184}$W 
  nuclei, respectively.} 
\label{figure_EDF}  
\end{figure}

The evaluation of the NME with the EDF approach was performed with the Gogny D1M functional~\cite{PRL_102_242501_2009}. 
In our EDF approach the various initial and final nuclear states are
given as superpositions of particle number and angular momentum projected Hartree-Fock-Bogoliubov states $|\Psi^{I}_{i/f}(\beta)\rangle$ with different quadrupole
deformation $\beta$ ($\beta<0$, $\beta=0$ and $\beta>0$ represent oblate, spherical and prolate shapes respectively)~\cite{PRL_105_252503_2010}: 
\begin{equation}
|I^{+,\sigma}_{i/f}\rangle=\sum_{\beta}g^{I\sigma}_{i/f}(\beta)|\Psi^{I}_{i/f}(\beta)\rangle,
\label{GCM_WF}
\end{equation}
where $g^{I\sigma}_{i/f}(\beta)$ are the coefficients of the linear combination which are obtained by solving the corresponding Hill-Wheeler-Griffin (HWG) equations and $\sigma$ labels the different states for a given angular momentum $I$~\cite{RingSchuck}. From these coefficients, the so-called collective wave functions - the amplitude of the states for a given deformation - can be extracted~\cite{RingSchuck}.
Fig.~\ref{figure_EDF} shows these amplitudes in the three lowest $0^+$ states in $^{184}$W and the
ground state in $^{184}$Os computed using the Gogny D1M functional. All these states are well prolate deformed. The two ground states have a similar distribution around $\beta\sim0.25$; the first excited $0^{+}$ state of $^{184}$W correspond to a vibration of the deformed ground state while the $0^{+}_{3}$ state has a larger prolate deformation $\beta'\sim0.35$. These amplitudes will affect significantly the final value of the NMEs as we will analyze below. 
In Table~\ref{tableEDF2} we show the excitation energy of the first excited 2$^+$ state $E(2_1^+)$ and its quadrupole transition probability $B(E2)$ to the ground state.  We find larger excitation energies of the first $2_{1}^{+}$ state and
reduced quadrupole transition probabilities compared to the experimental values. Tab.~\ref{tableEDF2} lists also the energies of the two lowest excited $0^+$ states in $^{184}$W. While the excitation energy of the $0^+_2$ state has a fair agreement with data, the $0^+_3$ excitation energy is too high, pointing to the need of including additional degrees of freedom. For example, the inclusion of pairing vibrations coupled to the quadrupole degree of freedom would lower the energies of the $0^{+}_{2},0^{+}_{3}$ states, bringing them closer to the experimental values. However, the quadrupole deformation of these states would not be very much affected~\cite{PLB_704_520_2011}. 
The NMEs from the $^{184}$Os ground state
to the three different $0^+$ states in $^{184}$W are obtained in this framework by folding
the deformation dependent wave functions with the Fermi and
Gamow-Teller matrix elements expressed in the deformed basis~\cite{PRL_105_252503_2010}: 
\begin{equation}
M^{\epsilon\epsilon}(0^{+}_{\sigma})=\langle0^{+,\sigma}_{^{184}\mathrm{W}}|\hat{M}_{F/GT}^{\epsilon\epsilon}|0^{+,\sigma=1}_{^{184}\mathrm{Os}}\rangle,
\end{equation}
where $\hat{M}^{\epsilon\epsilon}_{F/GT}$ are two-body Fermi and Gamow-Teller transition operators. The results are listed in Table~\ref{tableEDF2}.
We obtain that the Gamow-Teller component contributes about $75\%$ to the NME. For
comparison we have also listed the NME obtained if we hypothetically
assume spherical shapes for the initial and final nuclear states
(i.e. $\beta=\beta'=0$), obtaining values close to those of
ref.~\cite{Krivoruchenko2011140}. However, if the deformation of the
nuclear states is properly accounted for, the NMEs are reduced by about
an order of magnitude for the ground state transitions and even more
for the transitions to the excited states. To understand this behavior
we have plotted in Fig.~\ref{figure_EDF}(c) the Fermi component of
the NME as a function of the quadrupole deformations of the $^{184}$Os
($\beta$) and $^{184}$W ($\beta'$) nuclei (the Gamow-Teller
component shows the same general behavior):
\begin{eqnarray}
M_{F/GT}^{\epsilon\epsilon}(\beta,\beta')=\nonumber\\
\frac{\langle\Psi^{I=0}_{^{184}\mathrm{W}}(\beta')|\hat{M}_{F/GT}^{\epsilon\epsilon}|\Psi^{I=0}_{^{184}\mathrm{Os}}(\beta)\rangle}
{\sqrt{\langle\Psi^{I=0}_{^{184}\mathrm{W}}(\beta')|\Psi^{I=0}_{^{184}\mathrm{W}}(\beta')\rangle\langle\Psi^{I=0}_{^{184}\mathrm{Os}}(\beta)|\Psi^{I=0}_{^{184}\mathrm{Os}}(\beta)\rangle}}.
\end{eqnarray}
We observe two general trends: a) for
identical deformations of initial and final states
(i.e. $\beta=\beta'$) the NME decreases with increasing quadrupole
deformation; b) for fixed deformation $\beta$ of the initial state the
NME decreases with increasing difference between the deformations of
initial and final states. We find significant and similar quadrupole
deformations for both ground states, consistent with the large $B(E2)$
values observed experimentally. This explains why our NME values are
so much reduced compared to the assumed value in
ref.~\cite{Krivoruchenko2011140}. Although the first excited $0^+$
state in $^{184}$W has a similar deformation as the ground state, the
corresponding NME is smaller. This is caused by the fact that the wave
function of the first excited state has to be orthogonal to the one of
the ground state implying a nodal structure which leads to
cancellations when calculating the NME. In fact, we could interpret this state as a vibrational excitation of the ground state. The second excited $0^+$ state is
predicted by our calculation to have a noticeably larger deformation
than the ground state. This results in a smaller NME than for similar
deformations as obtained for the ground state due to the trends observed in
Fig.~\ref{figure_EDF}(c). Even if our calculation fails to predict the correct deformation of the $0_3^+$ state, the NME
to this state should be smaller than the one to the ground state. If its
deformation is different than the one of the ground state, the general
trend b) predicts a smaller NME. If the deformation is similar, there
should be strong cancellations in the NME caused by the orthogonality
to the two lower $0^+$ states. In this way, the NME for the capture to the ground state can
be interpreted as an upper limit.

The half-life of the resonant neutrinoless double-electron capture in $^{184}$Os is
related to Eq.~(\ref{EQ_1}) by $T_{1/2} = \ln
2/\lambda_{\epsilon\epsilon}$. Using $R = 5.382$ fm from
\cite{ADNDT_84_185_2004}, and $P_{\epsilon\epsilon} = 1.19617 \times
10^{33}$~eV$^6$ (in natural units) determined by the one-configuration Dirac-Fock method
for an extended nucleus~\cite{Droese:2012:Ref28} and the average of
the two values for $M^{\epsilon\epsilon}(0^+_3)$ in
Tab.~\ref{tableEDF2}, we obtain $T_{1/2} = 2.0\times 10^{30}$~y for
an effective neutrino mass of 1~eV. If we take the ground state NME
value as an upper limit (see above), we find a lower limit for the
half-life of $T_{1/2} = 1.3\times 10^{29}$~y. 


In conclusion, we have investigated the resonance condition and the
NME of double-electron capture in $^{184}$Os to three $0^+$
states in $^{184}$W. The $Q$ value was determined in a direct
measurement of the mass difference of $^{184}$Os to $^{184}$W to be
$1453.68(0.58)$ keV with a factor of 2.7 smaller uncertainty than
previously reported.  We have computed the NME in a microscopic
approach that consistently considers deformation degrees of freedom
and found that initial and final nuclear states are significantly
deformed. This deformation results in a suppression of the NME by a factor of
about 10 compared to a calculation that is restricted to the
spherical configuration. Combining our various improvements,
a lower limit of the half-life for the 0$\nu\epsilon\epsilon$ process
of $1.3 \times 10^{29}$~y to the excited $0^+$ state with 1322.152(22)
keV excitation energy in $^{184}$W, normalized to an effective
electron neutrino mass of 1 eV, has been obtained. Thus, $^{184}$Os can be excluded as a promising candidate for the search for neutrinoless double-electron capture.

We acknowledge helpful discussions with Fedor Simkovic. Financial
support by the Max-Planck Society, the Alliance Program of the
Helmholtz Association (HA216/EMMI), the Helmholtz International Center
for FAIR, and technical support by the Nuclear Chemistry
Department at the University of Mainz is acknowledged.  The work of
V.M.~Shabaev, I.I.~Tupitsyn, and N.A.~Zubova was supported by RFBR
(Grant No.10-02-00450). N.A.~Zubova acknowledges also support by the
``Dynasty'' foundation and by the FRRC.



%

\end{document}